\documentclass[12pt]{iopart}

\usepackage{hyperref}
\usepackage{graphicx}
\usepackage{caption}
\usepackage{subfig}
\usepackage[usenames]{color}
\usepackage{xspace}
\usepackage{multirow}
\usepackage{array}
\usepackage{booktabs}
\newcolumntype{C}[1]{>{\centering\let\newline\\\arraybackslash\hspace{0pt}}m{#1}}

\newcommand{\beq}{\begin{equation}}
\newcommand{\eeq}{\end{equation}}
\newcommand{\beqn}{\begin{eqnarray}}
\newcommand{\eeqn}{\end{eqnarray}}

\newcommand{\IGM}{{\texttt{IllinoisGRMHD}}\xspace}
\newcommand{\GiR}{{\texttt{GiRaFFE}}\xspace}
\newcommand{\GiRfood}{{\texttt{GiRaFFEfood}}\xspace}
\newcommand{\Kranc}{{\texttt{Kranc}}\xspace}
\newcommand{\OGM}{{\texttt{OrigGRMHD}}\xspace}

\def\eadnew#1#2{\address{#2 E-mail: \mailto{#1}}}

\begin{document}
\date{\today}
\title{GiRaFFE: An Open-Source General Relativistic Force-Free
  Electrodynamics Code}
\author{
  Zachariah B.~Etienne$^{1,2,*}$, 
  Mew-Bing Wan$^{3}$, 
  Maria C.~Babiuc$^{2,4}$, 
  Sean T.~McWilliams$^{2,5}$, 
  Ashok Choudhary$^{5}$}

\address{$^{1}$ Department of Mathematics, West Virginia University, Morgantown, WV 26506, USA}
\address{$^{2}$ Center for Gravitational Waves and Cosmology, West Virginia University, Chestnut Ridge Research Building, Morgantown, WV 26505, USA}
\address{$^{3}$ Institute for Advanced Physics and Mathematics, Zhejiang University of Technology, Hangzhou, 310032, China}
\address{$^{4}$ Department of Physics, Marshall University, Huntington, WV 25755, USA}
\address{$^{5}$ Department of Physics \& Astronomy, West Virginia University, Morgantown, WV 26506, USA}
\eadnew{zbetienne@mail.wvu.edu}{$^{*}$}

\begin{abstract}
We present \GiR, the first open-source general relativistic 
force-free electrodynamics (GRFFE) code for dynamical,
numerical-relativity generated spacetimes. \GiR adopts the strategy
pioneered by McKinney~\cite{McKinney:2006} and modified by Paschalidis
and Shapiro~\cite{Paschalidis:2013} to convert a GR
magnetohydrodynamic (GRMHD) code into a GRFFE code. In short, \GiR
exists as a modification of \IGM~\cite{Etienne:2015cea}, a
user-friendly, open-source, dynamical-spacetime GRMHD code. Both \GiR
and \IGM leverage the Einstein Toolkit's highly-scalable infrastructure
\cite{EinsteinToolkit,Carpet} to make possible large-scale simulations
of magnetized plasmas in strong, dynamical spacetimes on adaptive-mesh
refinement (AMR) grids. We demonstrate that \GiR passes a large suite
of both flat and curved-spacetime code tests passed by a number of
other state-of-the-art GRFFE codes, and is thus ready for
production-scale simulations of GRFFE phenomena of key interest to
relativistic astrophysics.
\end{abstract}


\maketitle

\section{Introduction}

With LIGO's first detections of gravitational
waves~\cite{Abbott:2016prlB,Abbott:2016prl}, the era of
gravitational wave (GW) astronomy has arrived, and an exciting new
window on the Universe has been opened. The prospect of
multimessenger observations stemming from an observed gravitational
wave source with coincident electromagnetic (EM) and/or neutrino
signals promises to provide far deeper insights into the extremely
violent processes that generate these GW signals.

0.4 seconds after
the first GW detection, a weak gamma-ray burst (GRB) lasting 1 second
was observed in the hard X-ray band ($>50$keV) by the Fermi Gamma-ray
Burst Monitor, in a region of the sky overlapping the very large LIGO
localization area~\cite{Connaughton:2016apjl}. Stellar-mass black hole
binaries (BHBs) are not generally expected to exist in matter-rich
environments assumed necessary to power such a GRB, suggesting that
the coincidence is most likely due to a statistical/systematic fluke in the
Fermi data, but would indicate the unexpected presence of a
tenuous plasma surrounding the BHB if the coincidence proved genuine
\cite{Belczynski:2002apj}. Regardless, if Nature does provide an
electromagnetic (EM) counterpart to the GW emission from 
BHBs---whether from a stellar-mass binary observable by Advanced LIGO,
intermediate-mass or massive BHBs observable by the Laser
Interferometer Space Antenna (LISA), or supermassive BHBs observable
by pulsar timing arrays---it will most likely be driven by the
interaction of the BHB with a tenuous plasma.

BHBs are not the only systems in which strong-field
gravitation could affect the dynamics of a tenuous,
magnetically-dominated plasma. Neutron star (NS) and pulsar
magnetospheres are other prime examples in which magnetic field lines
originating from the intense gravitational fields at the stellar
surface continue into the magnetosphere just outside the
NS. So just as in the case of BHBs, fully self-consistent modeling of
these systems requires that the plasma dynamics be followed in the
context of nonperturbative solutions to the general relativistic field
equations.

Future gravitational wave detections by LIGO may originate from binary
NSs or black hole--neutron star (BHNS) binary mergers. Observing an
electromagnetic counterpart to these incredibly violent processes may
provide deep insights not only into the behavior of matter at extreme
densities, but also the mechanism behind short GRBs. The behavior of binary
NS magnetospheres during their late-inspiral phase may launch EM
counterparts prior to and during a GW detection. In the case of BHNS
binaries, the motion of the BH through the magnetic field of its NS
companion during the inspiral can form a unipolar inductor, beaming
radiation over the full sky each orbit
\cite{Drell:1965,Goldreich:1969,McWilliams:2011,Paschalidis:2013jsa}.
At the end of the inspiral, either for a BHNS with a sufficiently
light, highly-spinning BH or a BNS with sufficiently high mass, NS
disruption and subsequent accretion onto the BH (formed from
hypermassive NS collapse in the case of BNS) can launch a classic
short GRB jet~\cite{Berger:2013jza,Nakar:2007yr,Paschalidis:2015,Paschalidis:2016,Ruiz:2016}.

In all of the aforementioned systems, the dynamics driving the EM
signal involve a strongly magnetized tenuous plasma accelerating
within a curved, often highly-dynamical spacetime. Properly modeling
such dynamics will require that the appropriate general relativistic
versions of the equations governing plasma dynamics be chosen.
For regions in which the magnetic-to-gas-pressure ratios are less than
or are of order unity, such as the interior of a NS or an accretion
disk surrounding a BH or BHB, the plasma dynamics are well-described
by the equations of ideal general relativistic magnetohydrodynamics
(GRMHD). But as these ratios far exceed unity---as is the case for
tenuous plasma outside BHs, BHBs, and NSs---the GRMHD equations
become stiff, and solving instead the equations of general
relativistic force-free electrodynamics (GRFFE) is far more
numerically tractable. The equations of GRFFE are a limiting case of
ideal GRMHD, in which the magnetic fields (as opposed to hydrodynamic
densities and pressures) entirely drive the plasma dynamics.

While technically it would be best to employ an algorithm that can
accommodate arbitrary magnetic-to-gas-pressure ratios, in practical
terms, the dynamics of many systems of interest are well approximated
by either the GRMHD or GRFFE limits. In exceptional cases such as the
boundary between a NS and its magnetosphere, an optimal method would
allow for feedback from the ideal GRMHD region into the GRFFE region,
and vice-versa. To this end, methods have been developed for feeding
information from the ideal GRMHD region of the NS interior into the
magnetosphere, which is well-described by the equations of GRFFE
(see \cite{Lehner:2011aa, Paschalidis:2013jsa, Paschalidis:2013}). 
  These methods effectively
impose a dynamical boundary condition for the surface of the
magnetosphere (i.e., the surface outside of which the equations of
GRFFE are solved). While GRMHD flows can inform regions well-described
by the GRFFE equations, the reverse is not yet possible with current
GRFFE prescriptions, as the GRFFE equations destroy information about
hydrodynamic pressures and densities---two essential quantities that
must be specified within the ideal GRMHD framework.

The GRFFE equations and their application in modeling important
astrophysical scenarios have a long history in the literature and
continue to be an active area of study by many independent groups.
Komissarov \cite{Komissarov:2002} was one of the first to develop a
conservative GRFFE formalism and numerical code in the context of
strong-field spacetimes to study the properties of the
plasma-filled magnetospheres of black holes \cite{Komissarov:2004,
  Komissarov:2005, Komissarov:2006}. 
Later, McKinney would introduce a general relativistic force-free
(GRFFE) numerical formulation that, through minimal modification of
a general relativistic magnetohydrodynamic (GRMHD) code, enables
evolutions of the GRFFE equations of motion
\cite{McKinney:2006, McKinney:2006a, McKinney:2004}.
Adopting such a prescription, Palenzuela~{\it et al}
\cite{Palenzuela:2011,Palenzuela:2010,Palenzuela:2010nf} studied the interaction of
binary and single black holes with ambient magnetic fields within the
GRFFE approximation, observing the formation of a dual jet morphology
when BHBs orbited within a uniform background magnetic field, and a
single jet in the context of a single, spinning black hole---a
dramatic confirmation of the Blandford-Znajek mechanism \cite{BZ:1977}
(see also~\cite{Spitkovsky:2006}).
Motivated in part by these results, as well as a desire to more deeply
explore important GRFFE phenomena in this age of multimessenger
astrophysics, multiple other independent groups have developed GRFFE
codes~\cite{Alic:2012, Cao:2015, Cao:2016, Nathanail:2014, Parfrey:2012a,
  Petri:2016mn, Petri:2016aa, Zhang:2015, Spitkovsky:2006}. 

One GRFFE code of particular interest to work presented here is that
of Paschalidis and Shapiro \cite{Paschalidis:2013}, who extended
McKinney's prescription for GRMHD-to-GRFFE code conversion. In part
this extension focused on smoothly matching general relativistic ideal
MHD to its force-free limit, so that GRMHD dynamics can inform regions
of the physical domain more accurately modeled by a GRFFE code. \IGM
is an open-source rewrite of the GRMHD code on which Paschalidis and
Shapiro's GRFFE code is based, and the code we present here,
\GiR\footnote{\GiR and both initial data modules (\GiRfood and {\tt ShiftedKerrSchild}) are
  open source (under the 2-clause BSD or GNU General Public License, version 2.0 or greater) and
  can be downloaded from \url{https://bitbucket.org/zach_etienne/wvuthorns/}.}, is
based on \IGM. Unlike the code of Paschalidis and Shapiro, \GiR is a
pure GRFFE code, not currently designed to match the equations of
GRMHD to its force-free limit.

\GiR represents the first open-source GRFFE code designed for
dynamical spacetime simulations. Based on \IGM, \GiR employs the
prescription devised by McKinney \cite{McKinney:2006} and refined by
Paschalidis and Shapiro~\cite{Paschalidis:2013} for converting a GRMHD
code into a GRFFE code. Like the code on which it was based, \GiR is
fully compatible with the Einstein Toolkit adaptive-mesh refinement
(AMR) Cactus/Carpet infrastructure \cite{EinsteinToolkit, Carpet}, and
is therefore highly scalable, to tens of thousands of cores. Also,
\GiR is quite compact, existing in only about 1,600 lines of code. For
dynamical-spacetime GRFFE simulations, \GiR may be immediately coupled
to McLachlan~\cite{Brown:2009McLachlan}, a state-of-the-art, open-source,
\Kranc-generated~\cite{Husa:2006,Kranc:web}
spacetime evolution module within the Einstein Toolkit.

The rest of the paper is structured as follows. In
Sec.~\ref{formalism}, we review our adopted GRFFE formalism and in
Sec.~\ref{algorithms} how these equations are solved within the \GiR
code. Then in Sec.~\ref{results}, we present a large suite of GRFFE
code tests, and finally in Sec.~\ref{conclusions} we summarize the
paper and present plans for future work.

\section{Adopted GRFFE Formalism}
\label{formalism}

All equations are written in geometrized units, such that
$G=c=1$, and Einstein notation is chosen for implied summation. Greek
indices span all 4 spacetime dimensions, and Latin indices span only
the 3 spatial dimensions.

The line element for our spacetime in standard 3+1 form is given by
\beq
ds^2 = -\alpha^2 dt^2 + \gamma_{ij} (dx^i + \beta^i dt)(dx^j + \beta^j dt),
\eeq
where $\alpha dt$ denotes the proper time interval between adjacent
spatial hypersurfaces separated by coordinate times $t=t_0$ and
$t=t_0+dt$, $\beta^i$ the magnitude of the spatial coordinate shift between adjacent
hypersurfaces, and $\gamma_{ij}$ is the three-metric within a given
hypersurface at coordinate time $t$.

Our 3+1 GRFFE formalism is written in terms of electric and magnetic
fields as measured by an observer co-moving and normal to the spatial
hypersurface, with 4-velocity $n^{\mu}=(1/\alpha,-\beta^i/\alpha)$. 
The stress-energy tensor of the electromagnetic field is defined as:
\beq
T_{\rm EM}^{\mu\nu} = F^{\mu}_{\sigma} F^{\nu \sigma} - \frac{1}{4}g^{\mu \nu}F_{\sigma \eta} F^{\sigma \eta}. 
\eeq
where $F^{\mu\nu}$ is the electromagnetic (or Faraday) tensor:
\beq
F^{\mu\nu} = n^{\mu} E^{\nu} - n^{\nu} E^{\mu} - \epsilon^{\mu\nu\sigma\eta}B_{\sigma}n_{\eta}, 
\eeq
In terms of the Faraday tensor, the electric $E^\nu$ and magnetic $B^\nu$ fields in this frame are given by
\beqn
E^{\nu} &=& n_{\mu} F^{\mu\nu} \\
B^{\nu} &=& -\frac{1}{2} n_{\mu} \epsilon^{\nu\mu\sigma\eta}
F_{\sigma\eta} =  n_{\mu} {}^*F^{\mu\nu},
\eeqn
where ${}^*F^{\mu\nu}$ is the dual of the Faraday tensor, 
and $\epsilon^{\nu\mu\sigma\eta}=[\nu\mu\sigma\eta]/\sqrt{|g|}$ 
is the rank-4 Levi-Civita tensor, with $[\nu\mu\sigma\eta]$ the
regular permutation symbol. In ideal MHD, the electric field vanishes
for observers moving {\it parallel} to the plasma with 4-velocity $u^\nu$:
\beq
F^{\mu\nu} u_{\nu} = -\sqrt{4 \pi} E^{\mu}_{(u)} = 0.
\eeq
I.e., when an observer moves in a direction parallel to the magnetic
field lines, the electric field vanishes.
This is simply a statement of Ohm's law for the case of perfect
conductivity. One implication of perfect conductivity is that
``magnetic field lines remain attached to the fluid elements they
connect''---the so-called ``frozen-in condition of MHD''. In addition,
so long as $F^{\mu\nu} \ne 0$, the ideal MHD condition implies that
\beqn
F^{\mu\nu} F_{\mu\nu} = 2(B^2-E^2) &>& 0\quad \rightarrow \quad B^2 > E^2,\quad {\rm and} \\
{}^*F^{\mu\nu} F_{\mu\nu} = 4 E_\mu B^\mu &=& 0. 
\eeqn
See, \textit{e.g.}, \cite{Komissarov:2002,Paschalidis:2013} for further discussion.

In addition to the ideal MHD condition $E^{\mu}_{(u)} = 0$, force-free
electrodynamics also assumes that the plasma dynamics are
completely driven by the electromagnetic fields (as opposed to, e.g.,
hydrodynamic pressure gradients).
This implies that the stress-energy of the plasma:
 $T^{\mu\nu}= T_{\rm matter}^{\mu\nu} + T_{\rm EM}^{\mu\nu}$,
is completely dominated by the electromagnetic
terms, which yields the conservation equation \cite{Palenzuela:2010,Parfrey:2012a}:
\beqn
&& \nabla_{\nu} T^{\mu\nu} \approx \nabla_{\nu} T^{\mu\nu}_{\rm EM}=-F^{\mu\nu} J_{\nu}=0 \\
\label{forcefree}
&& \rightarrow \rho E^i + n_\nu \epsilon^{\nu ijk} J^{(3)}_j B_k = 0.
\eeqn
where 
$J^{(3)}_i$ is
the 3-current, and $\rho$ the charge density. 
This formalism is valid in the tenuous plasma of the stellar magnetosphere, where the rest-mass density is vanishingly small and assumed to be zero. 
From these, the 4-current can be expressed 
$J^{\nu}=\rho n^{\nu}+\gamma^{i}_{j}J^{j}=\gamma^{\nu}_{\mu}J^{\mu}$ 

The left-hand side of Eq.~\ref{forcefree} is simply the general
relativistic expression for the Lorentz force, indicating that indeed
the Lorentz force is zero in GRFFE.
Notice that if we assume we are not in electrovacuum ($J^\mu \ne 0$),
multiplying Eq.~\ref{forcefree} by $B_i$ yields the familiar $B^i E_i
= 0$ constraint of force-free electrodynamics (see
\cite{Komissarov:2002,Paschalidis:2013} for full derivation).

In summary, the force-free constraints can be written
\beqn
\label{eq:BandEorthogconstraint}
&& B_i E^i = 0,\quad {\rm and}\\
&& B^2 > E^2.
\eeqn

Under these constraints, the GRFFE evolution equations
consist of the Cauchy momentum equation and the induction
equation (see
\cite{Komissarov:2002,McKinney:2006,Paschalidis:2013}
for derivation):
\begin{enumerate}
\item The Cauchy momentum equation follows directly from the spatial
  components of 
$\nabla_\mu T^{\mu\nu}=\nabla_\mu T^{\mu\nu}_{\rm  EM}=0$ (the time
  component yields the energy equation, which in GRFFE is
  redundant). We choose to write the momentum equation in conservative
  form and in terms of the densitized spatial Poynting flux one-form
  $\tilde{S}_i = \sqrt{\gamma}S_i$, 
\beq
\label{eq:momentum}
\partial_t \tilde{S}_i + \partial_j \left(\alpha \sqrt{\gamma}
T^j_{\rm EM}{}_i\right) = \frac{1}{2} \alpha \sqrt{\gamma}
T^{\mu\nu}_{\rm EM} \partial_i g_{\mu\nu},
\eeq
where $S_i$ can be derived from the expression of the Poynting one-form, 
\beq
\label{eq:Poynting}
S_{\mu}=-n_{\nu} T^{\nu}_{{\rm EM}\mu}.
\eeq

\item The induction equation in the force-free limit is derived from
  the spatial components of $\nabla_\mu {}^*F^{\mu\nu}=0$ (the time
  components yield the ``no-monopoles'' constraint), and can be
  written in terms of the densitized magnetic field $\tilde{B^i} =
  \sqrt{\gamma}B^i$ as
\beq
\label{eq:induction}
\partial_t \tilde{B}^i + \partial_j \left(v^j \tilde{B}^i - v^i \tilde{B}^j\right)=0,
\eeq
where $v^j = u^j/u^0$. 
As detailed in Appendix A of\cite{Paschalidis:2013}, the force-free
conditions do not uniquely constrain $u^\mu$, allowing for the freedom
to choose from a one-parameter family. As in
\cite{McKinney:2006,Paschalidis:2013}, we choose $u^\mu$ to correspond
to the {\it minimum} plasma 3-velocity that satisfies $F^{\mu\nu}
u_{\nu}=0$. This choice of $v^j$ is often referred to as the {\it
  drift} velocity, which can be defined in terms of known variables as
\beq
\label{eq:vfromS}
v^i = 4\pi \alpha \frac{\gamma^{ij} \tilde{S}_j}{\sqrt{\gamma}B^2}-\beta^i.
\eeq
\end{enumerate}

\section{Numerical Algorithms}
\label{algorithms}

We briefly review the numerical algorithms employed in \GiR to
solve the equations of GRFFE as outlined in Sec.~\ref{formalism}. 

\GiR fully supports Cartesian adaptive mesh refinement (AMR) grids via
the Cactus/Carpet \cite{Carpet} infrastructure within the Einstein
Toolkit \cite{EinsteinToolkit}.

As in \IGM, \GiR guarantees that the magnetic fields remain
divergenceless to roundoff error {\it even on AMR grids} by evolving
the vector potential $\mathcal{A}_\mu = \Phi n_\mu + A_\mu$, where
$A_\mu$ is spatial ($A_\mu n^\mu=0$), instead of the magnetic
fields directly. The vector potential fields exist on a staggered grid
(as defined in Table 1 of Ref.~\cite{Etienne:2015cea}) such that our
magnetic fields are evolved 
according to the flux constrained transport (FluxCT) algorithm of
Refs.~\cite{Balsara:1999,Toth:2000}.

Our choice of vector potential requires that we solve the vector
potential version of the induction equation 
\beq
\label{eq:Ainduction}
\partial_t A_i = \epsilon_{ijk} v^j B^k - \partial_i (\alpha \Phi-\beta^j A_j),
\eeq
where $\epsilon_{ijk} = [ijk] \sqrt{\gamma}$ is the anti-symmetric
Levi-Civita tensor and $\gamma$ is the 3-metric determinant, which in
a flat spacetime in Cartesian coordinates reduces to 1. $B^k$ in
Eq.~\ref{eq:Ainduction} is computed from the vector potential via
\beq
\label{eq:bfroma}
B^i=\epsilon^{ijk}\partial_j A_k=\frac{[ijk]}{\sqrt\gamma}\partial_j A_k.
\eeq

$\Phi$ is evolved via an additional electromagnetic gauge evolution
equation, which was devised specifically to avoid the buildup of
numerical errors due to zero-speed characteristic modes \cite{Etienne:2011re} on AMR grids. 
Our electromagnetic gauge is identical to the Lorenz gauge, but with an exponential
damping term with damping constant $\xi$ \cite{Farris:2012ux}:
\beq
\label{eq:EMgauge}
\partial_t \left[\sqrt{\gamma} \Phi \right] +
\partial_j \left(\alpha \sqrt{\gamma} A^j - \beta^j \left[\sqrt{\gamma} \Phi \right]\right)
= -\xi \alpha \left[\sqrt{\gamma}\Phi\right].
\eeq

{\bf Step 0: Initial data}: In addition to 3+1 metric quantities in
the Arnowitt-Deser-Misner (ADM) formalism \cite{Arnowitt:1959},
\GiR requires that the ``Valencia'' 3-velocity $\bar{v}^i$ and vector
potential one-form $A_\mu$ be set initially. Regarding the former, the
``Valencia'' 3-velocity $\bar{v}^i$ is related to the 3-velocity
appearing in the induction equation $v^i$ via 
\beq
v^i = \frac{u^i}{u^0} = \alpha \bar{v}^i-\beta^i.
\eeq
As for $A_\mu$, for all cases in this paper, we set the evolution
variable $\left[\sqrt{\gamma}\Phi\right]=0$ initially, and $A_i$ is
set based on our initial physical scenario.

After $v^i$ and $A_\mu$ are set, $B^i$ is computed via
Eq.~\ref{eq:bfroma}, and then the evolution variable $\tilde{S}_i$ is
given by
\beq
\label{eq:Sfromv}
\tilde{S}_i = \frac{\gamma_{ij} (v^j + \beta^j) \sqrt{\gamma} B^2}{4\pi\alpha}.
\eeq

{\bf Step 1: Evaluation of evolution equations}: In tandem with the
high-resolution shock-capturing (HRSC) scheme within \GiR, the
Runge-Kutta fourth-order (RK4) scheme is chosen to march
our evolution variables $A_i$ and $\tilde{S}_i$ forward in time from
their initial values, adopting precisely the same reconstruction and
Riemann solver algorithms as in \IGM (see Ref.~\cite{Etienne:2015cea}
for more details). In short, $A_i$ and $\tilde{S}_i$ are evolved
forward in time using the Piecewise Parabolic Method
(PPM)~\cite{Colella:1984} for reconstruction and a Harten-Lax-van Leer
(HLL)-based algorithm~\cite{Harten:1983,DelZanna:2003} for
approximately solving the Riemann problem. Meanwhile, spatial
derivatives within $\left[\sqrt{\gamma}\Phi\right]$'s evolution
equation (Eq.~\ref{eq:EMgauge}) are evaluated via finite difference
techniques (as in \IGM).

{\bf Step 2: Boundary conditions on $A_\mu$}: At the end of each RK4
substep, the evolved variables $A_i$ and $\tilde{S}_i$ have been
updated at all points except the outer boundaries. So next the
outer boundary conditions on $A_i$ and
$\left[\sqrt{\gamma}\Phi\right]$ are applied. As no exact outer
boundary conditions typically exist for systems of interest to \GiR,
we typically take advantage of AMR and push our outer boundary out of
causal contact from the physical system of interest. However, to retain
good numerical stability, we apply ``reasonable'' outer boundary
conditions. Specifically, values of $A_i$ and
$\left[\sqrt{\gamma}\Phi\right]$ in the interior grid are linearly
extrapolated to the outer boundary.

{\bf Step 3: Computing $B^i$}: $B^i$ is next computed from $A_i$ via
Eq.~\ref{eq:bfroma}.

{\bf Step 4: Applying GRFFE constraints \& computing $v^i$}:
Truncation, roundoff, and under-sampling errors will at times push
physical quantities into regions that violate the GRFFE
constraints. To nudge the variables back into a physically realistic
domain, we apply the same strategy as was devised in
Ref.~\cite{Paschalidis:2013} to guarantee that the GRFFE constraints
remain satisfied:

First, we adjust $\tilde{S}_i$ via
\beq
\tilde{S}_i \to \tilde{S}_i - \frac{(\tilde{S}_j \tilde{B}^j) \tilde{B}_i}{\tilde{B}^2}
\eeq
to enforce $B^i S_i=0$, which as shown by Ref.~\cite{Paschalidis:2013}, is
equivalent to the GRFFE constraint Eq.~\ref{eq:BandEorthogconstraint}.

Next, we limit the Lorentz factor of the plasma, typically to be
2,000, by rescaling $\tilde{S}_i$ according to Eq.~92 in
Ref.~\cite{Paschalidis:2013}. After $\tilde{S}_i$ is rescaled the
3-velocity $v^i$ is recomputed via Eq.~\ref{eq:vfromS}.

Finally, errors within our numerical scheme dissipate
sharp features, so when current sheets appear, they are quickly
and unphysically dissipated. This is unfortunate because current
sheets lie at the heart of many GRFFE phenomena. So to remedy the
situation, we apply the basic strategy of McKinney
\cite{McKinney:2006} (that was also adopted by Paschalidis and Shapiro 
\cite{Paschalidis:2013}) and set the velocity perpendicular to the
current sheet $v^\perp$ to zero. For example, if the current sheet
exists on the $z=0$ plane, then $v^\perp=v^z$, which we set to zero
via $n_i v^i=0$, where $n_i=\gamma_{ij} n^j$ is a unit normal one-form
with $n^j=\delta^{jz}$. Specifically, in the case of a current sheet
on the $z=0$ plane, we set
\beq
v^z = -\frac{\gamma_{xz} v^x + \gamma_{yz} v^y}{\gamma_{zz}}
\eeq
at all gridpoints that lie within $|z|\le 4\Delta z$ of the current
sheet.

At present the code addresses numerical dissipation of current
sheets only if they appear on the equatorial plane. For cases in which
current sheets appear off of the equatorial plane or spontaneously,
\cite{McKinney:2006} suggest the development of algorithms akin to
reconnection-capturing
strategies~\cite{StonePringle2001MNRAS.322..461S}. We intend to
explore such approaches in future work.

{\bf Step 5: Boundary conditions on $v^i$}: $v^i$ is set to zero at a
given face of our outermost AMR grid cube unless the velocity is {\it
  outgoing}. Otherwise the value for the velocity is simply copied
from the interior grid to the nearest neighbor on a face-by-face
basis.

After boundary conditions on $v^i$ are updated, all data needed for
the next RK4 substep have been generated, so we return to Step 1.

\begin{table}
\centering

\hspace*{-0.5cm}\begin{tabular}{C{3.0cm}C{7.7cm}C{5.6cm}} 
\hline\hline

{\footnotesize Test name \newline (wave speed)}& {\footnotesize Vector potential}& {\footnotesize Electric field} \\ \hline & \\[-2.2em]  

{\footnotesize Fast wave \newline ($\mu=1$)} &

{\footnotesize $\begin{array} {r@{}l@{}} 
& {} A_x = 0, \\ 
& {} A_y = 0, \\ 
& {} A_z = y+\left\{
\begin{array}{ll} -x-0.0075 & \mbox{if } x\leq -0.1, \\ 
0.75x^2-0.85x & \mbox{if } -0.1<x<0.1, \\ 
-0.7x-0.0075 & \mbox{if } x\geq 0.1. 
\end{array} \right. 
\end{array}$} &

{\footnotesize $\begin{array} {r@{}l@{}} 
& {} E^x = 0, \\ 
& {} E^y = 0, \\ 
& {} E^z = -B^y. 
\end{array}$} \\ \hline & \\[-2.2em]

{\footnotesize Alfv\'en wave \newline ($\mu=-0.5$)} &

{\footnotesize $\begin{array} {r@{}l@{}} 
& {} A_x = 0, \\ 
& {} A_y =\left\{
\begin{array}{ll} \gamma_{\mu}x-0.015 & \mbox{if } \gamma_\mu x\leq -0.1, \\ 
1.15\gamma_{\mu}x-0.03\cos(5\pi\gamma_{\mu}x)/\pi & \mbox{if } -0.1<\gamma_\mu x<0.1, \\ 
1.3\gamma_{\mu}x-0.015 & \mbox{if } \gamma_\mu x\geq 0.1, \\ 
\end{array} \right. \\ 
& {} A_z = y-\gamma_{\mu}(1-\mu)x. 
\end{array}$} &

{\footnotesize $\begin{array} {r@{}l@{}} 
& {} E^x = -B'^z, \\ 
& {} E^y = \gamma_{\mu}\mu B'^z, \\ 
& {} E^z = \gamma_{\mu}(1.0-\mu). 
\end{array}$} \\ \hline & \\[-2.2em]

{\footnotesize Degenerate Alfv\'en wave \newline ($\mu=0.5$)} &

{\footnotesize $\begin{array} {r@{}l@{}} 
& {} A_x = 0, \\ 
& {} A_y = \left\{
\begin{array}{ll} -0.8/\pi & \mbox{if } \gamma_{\mu} x\leq -0.1, \\ 
-(0.8/\pi)\cos[2.5\pi(\gamma_\mu x+0.1)] & \mbox{if } -0.1<\gamma_\mu x<0.1, \\ 
2(\gamma_{\mu}x-0.1) & \mbox{if } \gamma_\mu x\geq 0.1, 
\end{array} \right. \\ 
& {} A_z = \left\{
\begin{array}{ll} -2(\gamma_{\mu}x+0.1) & \mbox{if } \gamma_\mu x\leq -0.1, \\ 
-(0.8/\pi)\sin[2.5\pi(\gamma_\mu x+0.1)] & \mbox{if } -0.1<\gamma_\mu x<0.1, \\ 
-0.8/\pi & \mbox{if } \gamma_\mu x\geq 0.1. 
\end{array} \right. 
\end{array}$} &

{\footnotesize $\begin{array} {r@{}l@{}} 
& {} E^x = 0, \\ 
& {} E^y = \gamma_{\mu}\mu B'^z, \\ 
& {} E^z = -\gamma_{\mu}\mu B'^y. 
\end{array}$} \\ \hline & \\[-2.2em]

{\footnotesize Three waves \newline (stationary Alfv\'en: $\mu=0$, \newline fast right-going: \newline $\mu=1$, \newline fast left-going: \newline $\mu=-1$)} &

{\footnotesize $\begin{array} {r@{}l@{}} 
& {} A_x = 0, \\ 
& {} A_y = 3.5x\mathcal{H}(-x)+3.0x\mathcal{H}(x), \\ 
& {} A_z = y-1.5x\mathcal{H}(-x)-3.0x\mathcal{H}(x), \\ 
& {} \mbox{$\mathcal{H}$: Heaviside step function.} 
\end{array}$} &

{\footnotesize $\begin{array} {r@{}l@{}} 
& {} \mathbf{E} = \mathbf{E_a}+\mathbf{E_{+}}+\mathbf{E_{-}}, \\ 
& {} \mbox{stationary Alfv\'en wave:} \\ 
& {} \mathbf{E_a} =\left\{
\begin{array}{ll} (-1.0,1.0,0.0) & \mbox{if } x\leq 0, \\ 
(-1.5,1.0,0.0) & \mbox{if } x>0, \end{array} \right. \\ 
& {} \mbox{right-going fast wave:} \\ 
& {} \mathbf{E_{+}} =\left\{
\begin{array}{ll} (0.0,0.0,0.0) & \mbox{if } x\leq 0, \\ 
(0.0,1.0,-1.5) & \mbox{if } x>0, 
\end{array} \right. \\ 
& {} \mbox{left-going fast wave:} \\ 
& {} \mathbf{E_{-}} =\left\{
\begin{array}{ll} (0.0,-1.5,0.5) & \mbox{if } x\leq 0, \\ 
(0.0,0.0,0.0) & \mbox{if } x>0. 
\end{array} \right. 
\end{array}$} \\ \hline

{\footnotesize FFE breakdown \newline (none)} &

{\footnotesize $\begin{array} {r@{}l@{}} 
& {} A_x = 0, \\ 
& {} A_y = \left\{
\begin{array}{ll} x-0.2 & \mbox{if } x<0, \\ 
-5.0x^2+x+0.2 & \mbox{if } 0\leq x\leq 0.2, \\
-x & \mbox{if } x>0.2, 
\end{array} \right. \\ 
& {} A_z = y-A_y. 
\end{array}$} &

{\footnotesize $\begin{array} {r@{}l@{}} 
& {} \mathbf{E} = (0.0,0.5,-0.5). 
\end{array}$} \\ \hline\hline

\end{tabular}\hspace*{-0.2cm}
\caption{Initial setup for 1D tests. The unprimed and primed fields
  are in the grid and wave frames respectively, evaluated as functions
  of the coordinate $x$. $\mu$ is the wave speed relative to the grid
  frame (in units where $c=1$), and $\gamma_{\mu}$ is the
  corresponding relativistic Lorentz factor. The fields in the wave
  frame can be obtained from that in the grid frame via a simple
  Lorentz boost (see Eqs.~104 and 105 in
  Ref.~\cite{Paschalidis:2013}). From the $A_i$ and $E_i$ given above,
  all other quantities $v^i$, $B^i$ and $\tilde{S}_i$ needed
  for a GRFFE evolution can be computed from Eqs.~\ref{eq:Poynting},\ref{eq:vfromS},
  and \ref{eq:bfroma} respectively.} 
\label{tab:1Dinitdata}
\end{table}

\begin{table} 
\centering
\begin{tabular}{C{2.5cm}C{2.5cm}C{2.5cm}C{2.5cm}C{1.8cm}}
\hline\hline

Test name & 
$(N_x,N_y,N_z)$ & 
$[x_{min},x_{max}]$ & 
$[y_{min},y_{max}]$ \&  
$[z_{min},z_{max}]$ & 
CFL factor\\ \hline & \\[-2.0em]

Fast wave &  

\multirow{4}{*}{$\begin{array} {r@{}l@{}} 
(1280,32,32) 
\end{array}$} & \quad 

\multirow{4}{*}{$[-4.0,4.0]$} & 
\multirow{4}{*}{$[-0.025,0.025]$} \\[-0.7em] \cline{1-1}  

Alfv\'en wave \\ \cline{1-1} 
Degen. Alfv\'en \\ \cline{1-1} 
Three waves & & & & 

\multirow{2}{*}{$0.5$} \\ \cline{1-4}  

FFE breakdown &  

{$\begin{array} {r@{}l@{}} 
(200,8,8) 
\end{array}$} & 

$[-0.4,0.6]$ & 

\multicolumn{1}{c}{$\begin{array} {r@{}l@{}} 
[-0.02,0.02] 
\end{array}$} \\ \hline\hline

\end{tabular}
\caption{Grid setup for 1D tests.} 
\label{tab:1Dgrid}
\end{table}

\begin{table}
\centering
\small
\renewcommand{\arraystretch}{1.4}

\begin{tabular}{C{2.7cm}C{4.4cm}C{7.3cm}}
\hline\hline

Test name & Vector potential & Electric field \\
\hline

Split Monopole & 

$\begin{array} {r@{}l@{}} 
A_r & {} = -\frac{\displaystyle Ca}{\displaystyle 8}|\cos\theta|\times \\ 
& {} \quad\left( 1+\frac{\displaystyle 4M}{\displaystyle r} \right) \sqrt{1+\frac{\displaystyle 2M}{\displaystyle r}}, \\[2pt] 
A_{\phi} & {} = C M^2 \left(1-|\cos\theta|+\right. \\
& {} \left.\quad\quad a^2 f(r)\cos\theta\sin^2\theta\right). 
\end{array}$ &

$\begin{array} {r@{}l@{}} 
E_r & {}= -\frac{\displaystyle Ca^3}{\displaystyle 8\alpha M^3}f'(r)\cos\theta\sin^2\theta, \\[2pt] 
E_{\theta} & {}= -\frac{\displaystyle Ca}{\displaystyle 8\alpha}
\left[\sin\theta+\right. \\
& {} \left.\quad\quad a^2f(r)\sin\theta(2\cos^2\theta-\sin^2\theta)\right]- \\[2pt] 
& {} \quad\quad \beta^r\sqrt\gamma\frac{\displaystyle Ca}{\displaystyle 8r^2}\left( 1+\frac{\displaystyle 4M}{\displaystyle r} \right), \\[2pt] 
E_{\phi} & {}= \frac{\displaystyle \beta^r}{\displaystyle \alpha M}Ca^2f'(r)\cos\theta\sin^2\theta, \\[2pt] 
f(r) & {}= \frac{\displaystyle r^2(2r-3M)}{\displaystyle 8M^3}L\left( \frac{\displaystyle 2M}{\displaystyle r} \right) \\[2pt] 
& {} \quad +\frac{\displaystyle M^2+3Mr-6r^2}{\displaystyle 12M^2}\ln\frac{\displaystyle r}{\displaystyle 2M} \\[2pt] 
& {} \quad +\frac{\displaystyle 11}{\displaystyle 72}+\frac{\displaystyle M}{\displaystyle 3r}+\frac{\displaystyle r}{\displaystyle 2M}-\frac{\displaystyle r^2}{\displaystyle 2M^2}, \\[2pt] 
L(x) & {}= \mbox{Li}_2(x)+ \\
& {} \quad \frac{\displaystyle 1}{\displaystyle 2}\ln x\ln(1-x) \quad \mbox{for } 0<x<1. 
\end{array}$ \\[1ex] \hline 

Exact Wald & 

$A_{\phi}=\frac{\displaystyle C_0}{\displaystyle 2}r^2\sin^2\theta.$ &

$\begin{array} {r@{}l@{}} 
& {} E_r = 0, \\ 
& {} E_{\theta} = 0, \\ 
& {} E_{\phi}=2MC_0\left( 1+\frac{\displaystyle 2M}{\displaystyle r} \right)^{-1/2}\sin^2\theta. 
\end{array}$ \\[1ex] \hline

Magnetospheric Wald & 

$A_i=\frac{\displaystyle C_0}{\displaystyle 2}(g_{i\phi}+2ag_{ti}).$ & 

$\mathbf{E}=0.$ \\[1ex] \hline\hline

\end{tabular}

\caption{Initial setup for 3D BH spacetime tests. 
  $a$ is the dimensionless spin parameter for the BH with $M$ representing the BH mass.
  $C$ and $C_0$ are constants (both set to 1 in our simulations), and $L(x)$ is the
  dilogarithm function. To avoid the logarithmic divergence as
  $r\to\infty$, we neglect $f(r)$ and $f'(r)$ terms in our tests,
  setting them to zero. 
  From the $A_i$ and $E_i$ given above, all other quantities $v^i$, $B^i$ and $\tilde{S}_i$ 
  needed for a GRFFE evolution can be computed from Eqs.~\ref{eq:Poynting},
  \ref{eq:vfromS}, and \ref{eq:bfroma} respectively.}
\label{tab:3Dinitdata}
\end{table}

\begin{table}  
\centering
\begin{tabular}{C{3cm}C{5.7cm}C{5cm}}
\hline\hline

Test name & 
{$\begin{array} {r@{}l@{}} 
& {} \mbox{AMR cube half-side length;} \\
& {} \mbox{\quad\quad\quad\quad Resolution}
\end{array}$} & 
CFL factor \\ \hline 
 
Split Monopole &  

{$\begin{array} {r@{}l@{}}
& {} 3.125\times 2^{6-n}M; \\
& {} \frac{M}{16}\times 2^{6-n},\quad n=1,2,...,6 
\end{array}$} & \\[5ex] \cline{1-2} 

Exact Wald & 

\vspace*{-3mm}
{$\begin{array} {r@{}l@{}}
& {} 3.125\times 2^{6-n}M; \\
& {} \Delta x_{\rm min}\times 2^{6-n},\quad n=1,2,...,6, \\
& {} \Delta x_{\rm min}=\frac{M}{12.8}\mbox{(L)},\frac{M}{25.6}\mbox{(M)},\frac{M}{51.2}\mbox{(H)} 
\end{array}$} \vspace{0.5mm} &  

{$\begin{array} {r@{}l@{}}
& {} \frac{1}{32}\mbox{ for }n=1, \frac{1}{16}\mbox{ for }n=2, \\
& {} \frac{1}{8} \mbox { for } n=3,4,5,6
\end{array}$} \\ \cline{1-2}

Magnetospheric Wald & 

{$\begin{array} {r@{}l@{}}
& {} 3.125\times 2^{6-n}M; \\
& {} \frac{M}{8}\times 2^{6-n},\quad n=1,2,...,6 
\end{array}$} & \\[5ex] \hline 

& \\[-2.2em]
Aligned Rotator & 

$\begin{array} {r@{}l@{}}
& {} 2.94R_{\rm NS}\times 2^{10-n}; \\
& {} \Delta x_{\rm min}\times 2^{10-n},\quad n=1,2,...,10, \\
& {} \Delta x_{\rm min}=0.0147 R_{\rm NS}
\end{array}$ & 

$\begin{array} {r@{}l@{}}
& {} \frac{1}{20} \times 2^{n-1},n=1,2,3, \\
& {} \frac{2}{5} \mbox { for } n=4,...,10
\end{array}$ \\ \hline

\end{tabular}
\caption{Grid setup for 3D tests. $n=1$ and $\Delta
  x_{\rm min}$ represent the coarsest refinement
  level, and the grid spacing of the finest level respectively. L, M,
  H denote low, medium, and high resolution respectively. The BH mass
  $M$ is set to 1 in all tests for numerical convenience
  (all results are scale-free and can be trivially rescaled to any
  desired mass), and $R_{\rm NS}=1$ for the Aligned
  Rotator test, also for numerical convenience due to the case being
  scale-free according to $R_{\rm NS}$.}
\label{tab:3Dgrid}
\end{table}

\section{Results}
\label{results}

We focus on the same suite of flat and curved spacetime background code
validation tests as in Paschalidis {\it et
  al.}~\cite{Paschalidis:2013}, who in turn take cues from code
validation tests of Komissarov~\cite{Komissarov:2002} and McKinney~\cite{McKinney:2006}. 
The tests are designed to push GRFFE codes to their limits, while at the same time providing
either an exact or well-described qualitative solution to which
a code can be demonstrated to converge with increasing numerical
resolution.
 
The GRFFE code described in Ref.~\cite{Paschalidis:2013} exists as a
modification to \OGM, the original GRMHD code of the Illinois Numerical
Relativity group. \IGM is a complete open-source rewrite of \OGM,
demonstrated to generate results that agree to roundoff error with
\OGM. \GiR is based on a modification of \IGM, so we expect that
the results of the tests we perform here would match closely with
those presented in Ref.~\cite{Paschalidis:2013}. However, we are
unable to demonstrate roundoff-level agreement between \GiR and the
code of Ref.~\cite{Paschalidis:2013} since we do not have access to
the latter. So wherever possible, we attempt to duplicate figures
presented in Ref.~\cite{Paschalidis:2013} with \GiR so as to make
direct comparison straightforward. In short, we find excellent
qualitative agreement between \GiR and the results of
Ref.~\cite{Paschalidis:2013}.

The rest of this section is organized as
follows. Section~\ref{ssec:flatspacetimeresults} presents our
flat-spacetime code validation results, and
Sec.~\ref{ssec:curvedspacetimeresults} details curved-spacetime
results.

\subsection{Flat spacetime background tests} \label{ssec:flatspacetimeresults}

Our flat spacetime tests include a suite of five tests in one spatial
dimension (i.e., our ``1D code tests''), as well as the aligned
rotator test, which is a toy model of a pulsar magnetosphere. We
largely follow the testing procedures outlined in
Refs.~\cite{Komissarov:2002,McKinney:2006,Paschalidis:2013}.

Initial data parameters for the 1D tests are summarized in
Table~\ref{tab:1Dinitdata} and in Sec.~\ref{sssec:aligrot} for the
aligned rotator. Numerical grid parameters are provided in
Table~\ref{tab:1Dgrid} for the 1D tests and in Table~\ref{tab:3Dgrid}
for the aligned rotator. So that a direct comparison of the test
results can be made between our results and those presented
in~\cite{Paschalidis:2013}, our numerical grids and initial data
parameters are chosen to match those in~\cite{Paschalidis:2013}.

1D code test results are presented in Fig.~\ref{fig:1Dtests}. The
solutions for the tests at time $t$ are denoted as 
$Q(t,x)=Q(0,x+\mu t)$, where $Q$ represents physical quantities $B^i$,
$E^i$, or $v^i$, and $\mu$ is the wave velocity (in geometrized units
with $c=1$).

\subsubsection{Fast Wave}

The fast wave, either right or left-going, is one of the
characteristic waves of an FFE system. This test for a right-going
fast wave is based on that originally done in
Ref.~\cite{Komissarov:2002} for an FFE system. Since the fast wave
propagates at the speed of light, $\mu=1$. The analytic solution at
time $t=0.5$ is obtained by shifting the wave at the initial time to
the right by $x\to x+0.5$. Comparing the $z$-component of the electric
field with the numerical solution (top panel of Fig.~\ref{fig:1Dtests}),
we see a complete overlap, indicating that our numerical result
matches the analytic solution at this high resolution.

\subsubsection{Alfv\'en Wave}

The right and left-going Alfv\'en waves are also characteristic waves
of an FFE system, and similar to the fast wave, we base this test on
the left-going Alfv\'en wave test originally performed in
Ref.~\cite{Komissarov:2002}. For the analytic solution, we shift the
wave at the initial time by $x\to x-1$, and compare this with the
numerical solution at time $t=2$. The extremely close overlap between
the analytic and numerical solutions, as evidenced in the
$z$-component of the magnetic field, shown in the second-from-top
panel of Fig.~\ref{fig:1Dtests}, indicates that \GiR reproduces
the analytic solution quite well.

\subsubsection{Degenerate Alfv\'en Wave}

This test is originally performed in Ref.~\cite{Komissarov:2002} to
evolve a system in which the right and left-going Alfv\'en waves
possess the same wave speed. The degenerate Alfv\'en wave speed is
given as \cite{Komissarov:2002,Paschalidis:2013}:
\beq
\label{eq:degalfwavemu}
\mu=\frac{B_z E_y-B_y E_z}{B^2},
\eeq
consistent with the velocity computed via Eq.~\ref{eq:vfromS}. To
compare the analytic and numerical solutions at time $t=1$, we shift
the wave at the initial time by $x\to x+0.5$. For the $y$-component of
the electric field, we see a very good agreement (middle panel of
Fig.~\ref{fig:1Dtests}) between numerical and analytical results.

\subsubsection{Three Waves}

The three waves test initial data are constructed by superposing the
stationary Alfv\'en wave with the right and left-going fast
waves. This test was originally performed in
Ref.~\cite{Komissarov:2002}. For the comparison between the analytic
and numerical solutions at time $t=0.5625$, we shift the initial
left-going fast wave by $x=-0.5625$ and the initial right-going fast
wave by $x=+0.5625$, while maintaining the position of the initial
stationary Alfv\'en wave. The second-from-bottom panel of
Fig.~\ref{fig:1Dtests} demonstrates an almost complete overlap between
the analytic and numerical solutions, as expected for this high resolution.

\subsubsection{FFE breakdown} 

\begin{figure}[p]
\centering
\includegraphics[angle=0,width=\textwidth,height=0.9\textheight,keepaspectratio=true]{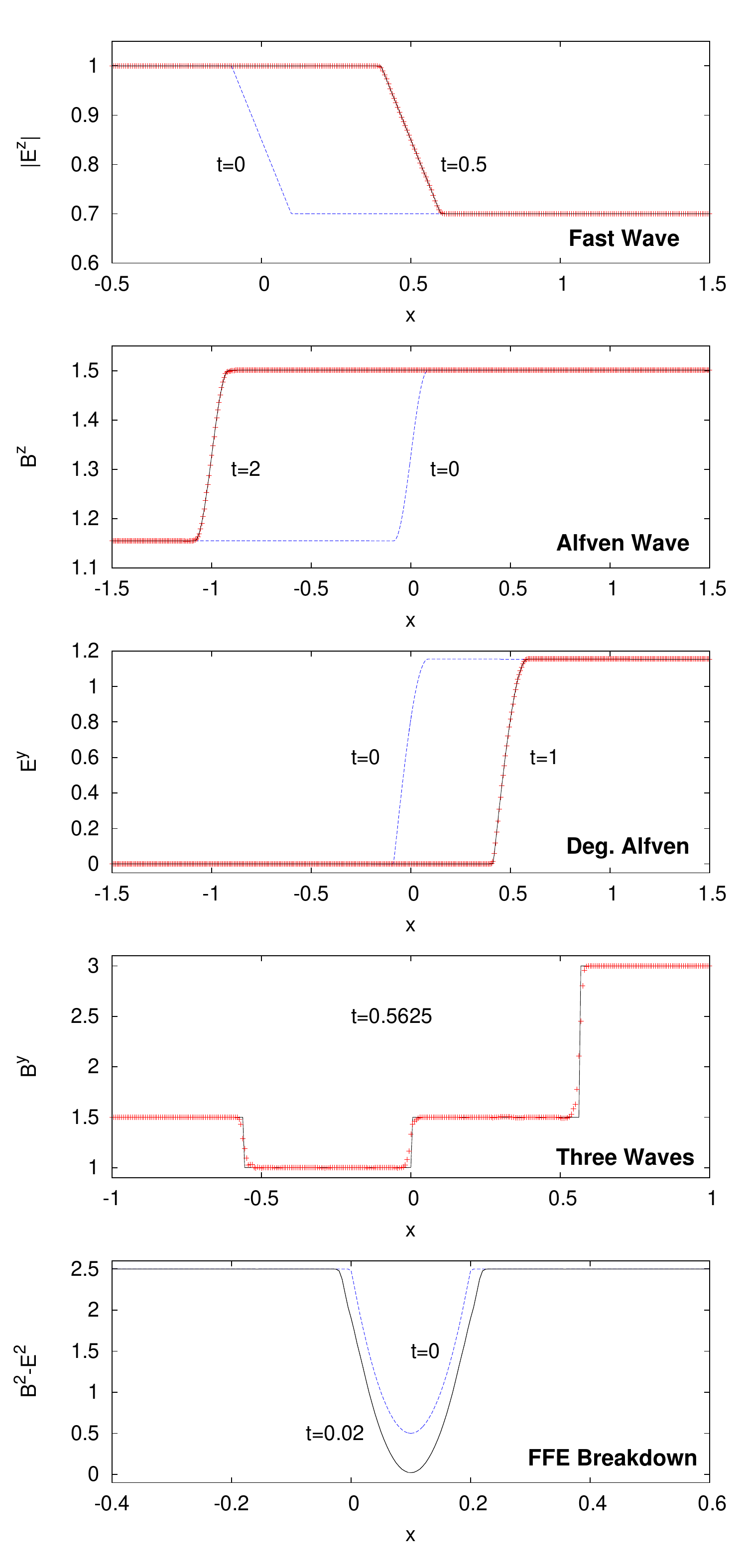}
\caption{Summary of 1D test results. In the top four panels,
  initial data are shown as dashed blue lines. At later times, the
  analytic solution and the numerical solution are shown as solid
  black lines and red crosses respectively. The bottom panel shows the
  numerical solution as a solid black line. 
  Figure formatting and numerical grid resolutions duplicate that of
  Ref.~\cite{Paschalidis:2013} so that direct comparisons can be made.}
\label{fig:1Dtests}
\end{figure}

The FFE breakdown test is originally performed in
Ref.~\cite{Komissarov:2002} to show that systems with left and right
states satisfying the FFE conditions may violate these conditions
numerically as time progresses. The initial data consists of a
transition layer which obeys the FFE conditions, but evolves to a
state where it does not \cite{Komissarov:2002}. Fig.~\ref{fig:1Dtests}
shows that $B^2-E^2$ decreases in time and at $t\approx 0.02$,
$B^2-E^2$ approaches zero, signaling the breakdown of the FFE
condition \cite{Komissarov:2002}. Our results match quite closely to
those of Ref.~\cite{Paschalidis:2013}.

\subsubsection{Aligned Rotator} \label{sssec:aligrot}

This test is based on that performed
in Refs.~\cite{Komissarov:2006,McKinney:2006a,Spitkovsky:2006,Paschalidis:2013}
to study a time-dependent toy model of a pulsar magnetosphere. The
test consists of a spherical surface (the ``surface of the star'')
that rotates at constant angular velocity, with an initially {\it
  stationary} dipolar magnetic field that threads the surface and
extends to $r\to\infty$ (i.e., the outer boundary in our numerical
simulations).

Specifically, the initial magnetic field is set via the vector potential
\beq
\label{eq:aligrotpot}
A_{\phi}=\frac{\mu\varpi^2}{r^3};\ \ A_r=A_\theta=0,
\eeq
where $\mu=B_p R_{\rm NS}^3/2$, $R_{\rm NS}$ is the stellar radius, and
$\varpi=\sqrt{x^2+y^2}$ the cylindrical radius. The velocity field at
the surface and inside the ``star'' is fixed for all times to be a
solid-body rotator:
\beq
\mathbf{v}=\Omega\mathbf{e}_z\times\mathbf{r},
\eeq
where $\mathbf{e}_z$ is the unit vector in the $z$-direction, and
$\Omega$ is the angular velocity of the star.

We set $\Omega=0.2$ such that the light cylinder is located at $R_{\rm LC}=5 R_{\rm NS}$, as in
\cite{Paschalidis:2013}. Inside the star, the evolution of the
densitized Poynting flux is disabled to ensure the interior velocity
fields maintain solid-body rotation, but the $A_i$ and
$\left[\sqrt{\gamma} \Phi \right]$ fields are evolved so as to
avoid kinks in the magnetic field from appearing at the stellar
surface.

Since the solution in this test involves a current sheet within the
light cylinder on the equatorial plane, i.e., on the $z=0$ plane, we
apply our strategy for handling current sheets as outlined in Step 4
of \GiR's core numerical algorithm (Sec.~\ref{algorithms}).

\begin{figure*}
\centering
\subfloat[]{%
\includegraphics[angle=0,scale=0.4]{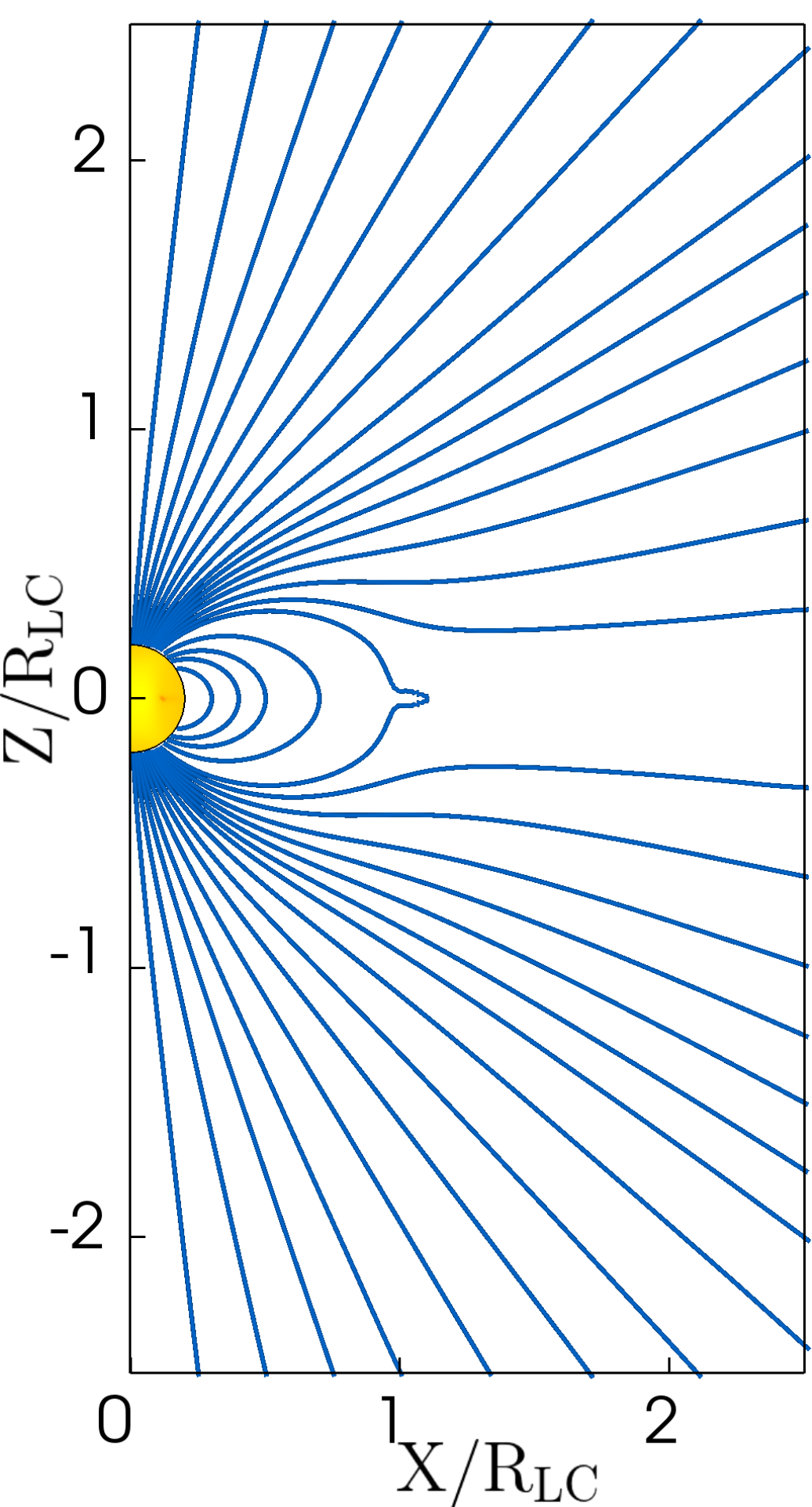}%
}%
\quad
\subfloat[]{%
\includegraphics[angle=0,scale=0.4]{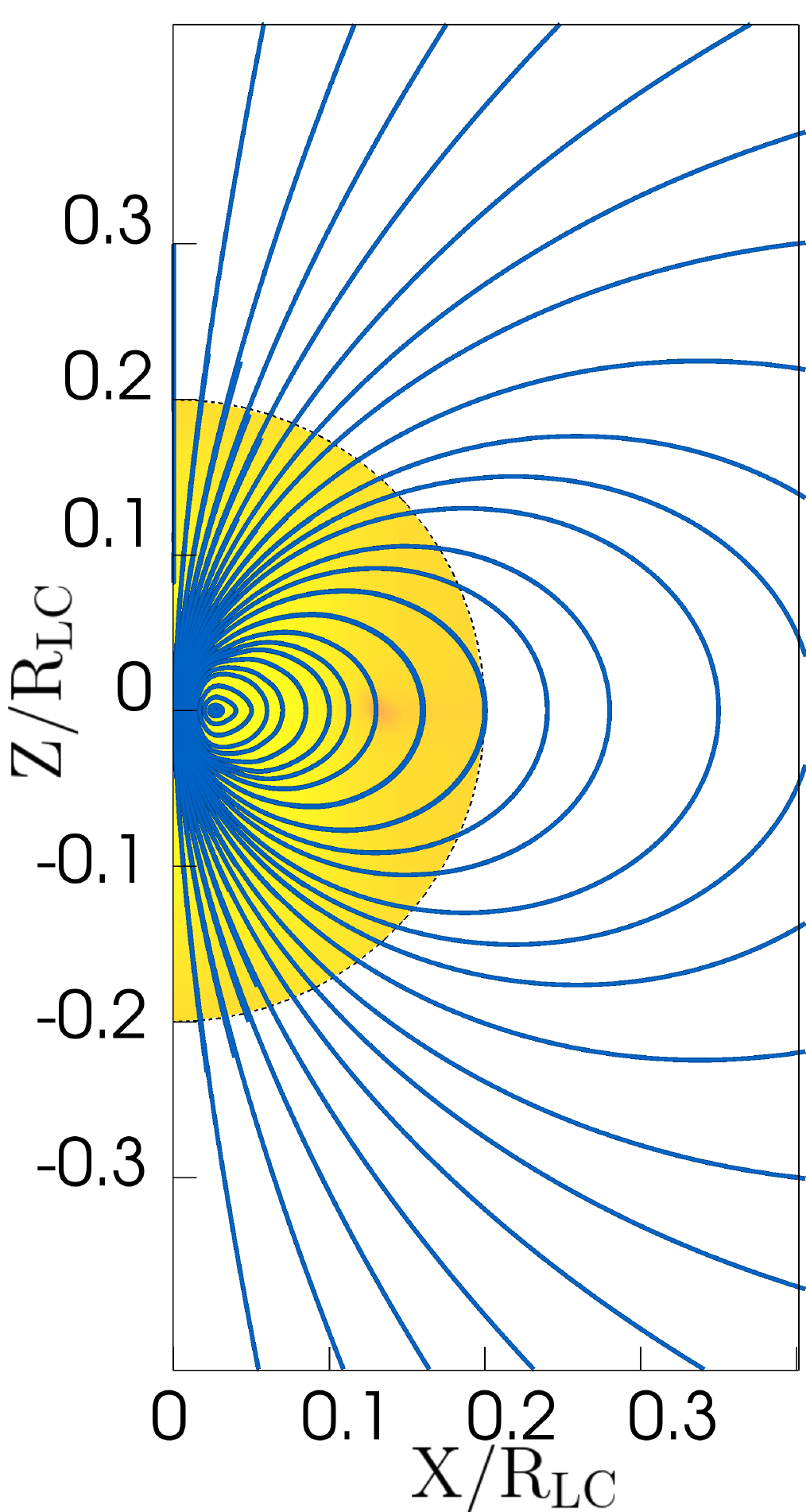}%
}%
\quad
\subfloat[]{%
\includegraphics[angle=0,scale=0.3]{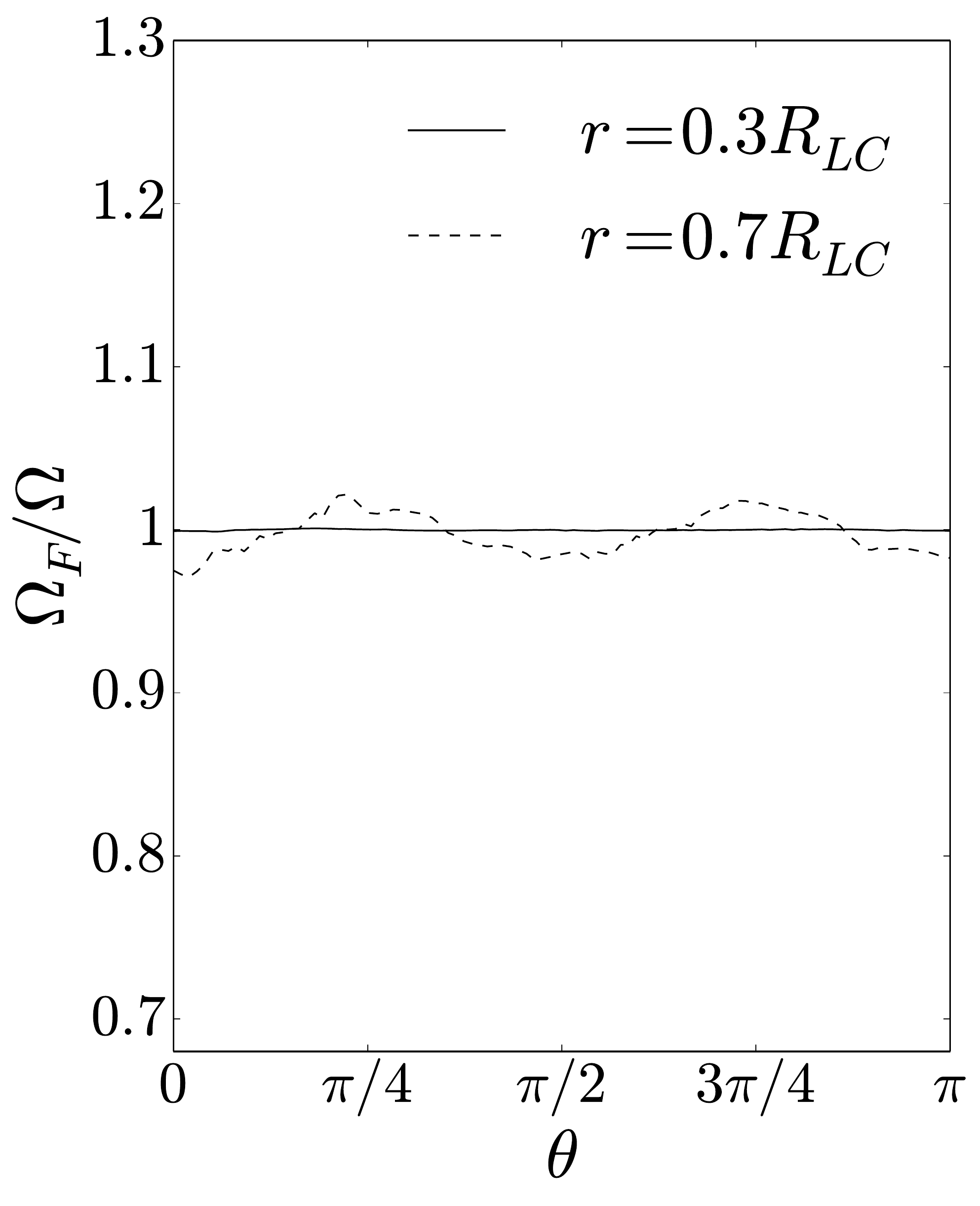}%
}%
\caption{%
  Aligned Rotator: a) Magnetic field structure after 3 rotation
  periods of the rotator. $R_{\rm LC}$ denotes light-cylinder
  radius. b) Magnetic field structure near the ``star'' (yellow circle; zoomed out in a)). 
  c) Plasma orbital frequency at 30\% and 70\%
  the light-cylinder radius, versus the azimuthal angle $\theta$.
  Figure formatting duplicates that of Ref.~\cite{Paschalidis:2013}
  so that direct comparison can be made, and we note that our grid
  structure corresponds most closely to their ``low-resolution''
  case.}
\label{fig:AlignedRotatorBfieldsandOmega}%
\end{figure*}

After starting the simulation, \GiR quickly reproduces known features
from the stationary solution of Contopoulos \textit{et al}
\cite{Contopoulos:1999}. These features were later corroborated via
the MHD simulation of Komissarov \cite{Komissarov:2006} and the FFE
simulation of McKinney \cite{McKinney:2006a}. In short, the surface of
the solid-body rotation of the stellar surface spins up the magnetic
field lines in the magnetosphere to the same angular frequency as the
stellar surface. In Fig.~\ref{fig:AlignedRotatorBfieldsandOmega}, we
show that the angular frequency of the magnetic fields closely tracks
the angular velocity of the star, even out to $r=0.7R_{\rm LC}$ {\it
  regardless of the angle at which $\Omega$ is measured at this
  radius}. And as $r\to R_{\rm LC}$, the magnetic field lines
transition from dipolar in structure (left panel of
Fig.~\ref{fig:AlignedRotatorBfieldsandOmega}) to an open field line
configuration (center panel of
Fig.~\ref{fig:AlignedRotatorBfieldsandOmega}).

\subsection{Curved spacetime background tests} \label{ssec:curvedspacetimeresults}

We perform a set of three curved-spacetime background tests: the split
monopole, the exact Wald solution, and the magnetospheric Wald solution.
Initial data parameters are summarized in Table~\ref{tab:3Dinitdata}, and
the grid setups are presented in Table~\ref{tab:3Dgrid}.
Testing procedures largely follow that outlined in Ref.~\cite{Paschalidis:2013},
in which plasma dynamics are modeled near a BH in ``shifted
Kerr-Schild'' coordinates (i.e., Kerr-Schild coordinates but with the
radial coordinate shifted to minimize the strong curvature near $r=0$;
see Appendix for full 3+1 decomposition). The equatorial
current sheet plays an important role in these tests, and to prevent
numerical dissipation intrinsic to our algorithms from quickly
destroying the current sheet, the strategy of Ref.\cite{McKinney:2006} is
employed as described in Sec.~\ref{algorithms}.

\subsubsection{Split Monopole}

The split monopole solution is derived from the Blandford-Znajek
force-free monopole solution~\cite{BZ:1977, McKinney:2004}, by
inverting the solution in the lower hemisphere. The solution we use in
the test is accurate only to first order in $a$ and follows
Refs.~\cite{Komissarov:2004,McKinney:2006,Paschalidis:2013} in dropping
problematic terms involving $f(r)$ and $f^\prime(r)$. The test is
performed in shifted Kerr-Schild radial coordinates (see Appendix)
with radial shift $r_0=1.0 M$, and with spin $a_* = a/M=J/M^2 = 0.1$
where $M$ is the BH mass. One important property of this solution is
that although initially all magnetic field lines penetrate the black
hole horizon, later they escape from the ergosphere of the black
hole. Therefore, a stable equatorial current sheet is required to
sustain this configuration, otherwise the magnetic field lines
reconnect and are pushed away~\cite{Komissarov:2004}. 

\begin{figure*}
\centering
\subfloat[]{%
\includegraphics[angle=-90,scale=0.45]{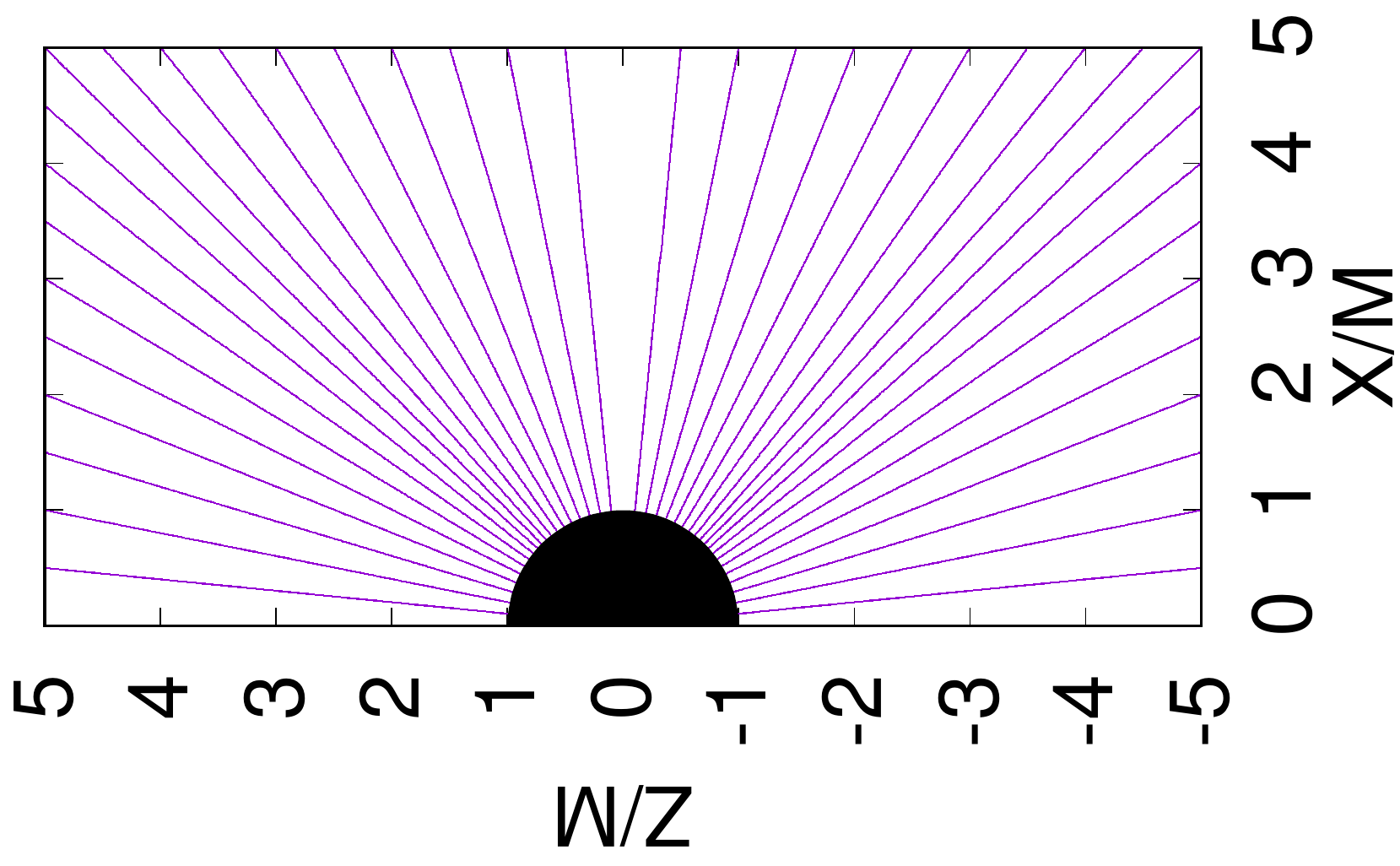}%
}%
\quad
\subfloat[]{%
\includegraphics[angle=-90,scale=0.45]{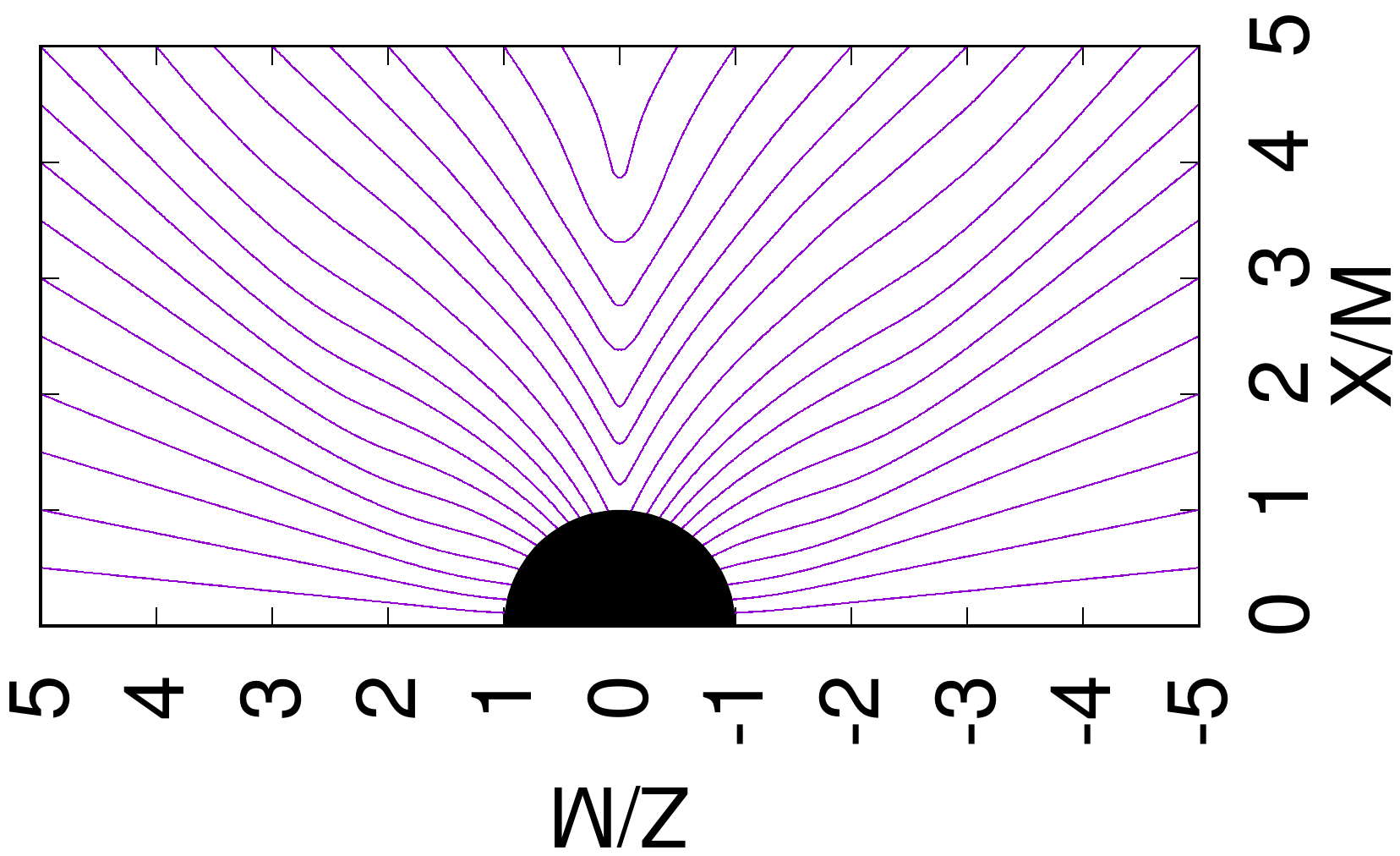}%
\label{fig:SplitMonopoleb}
}%
\quad
\subfloat[]{%
\includegraphics[angle=-90,scale=0.45]{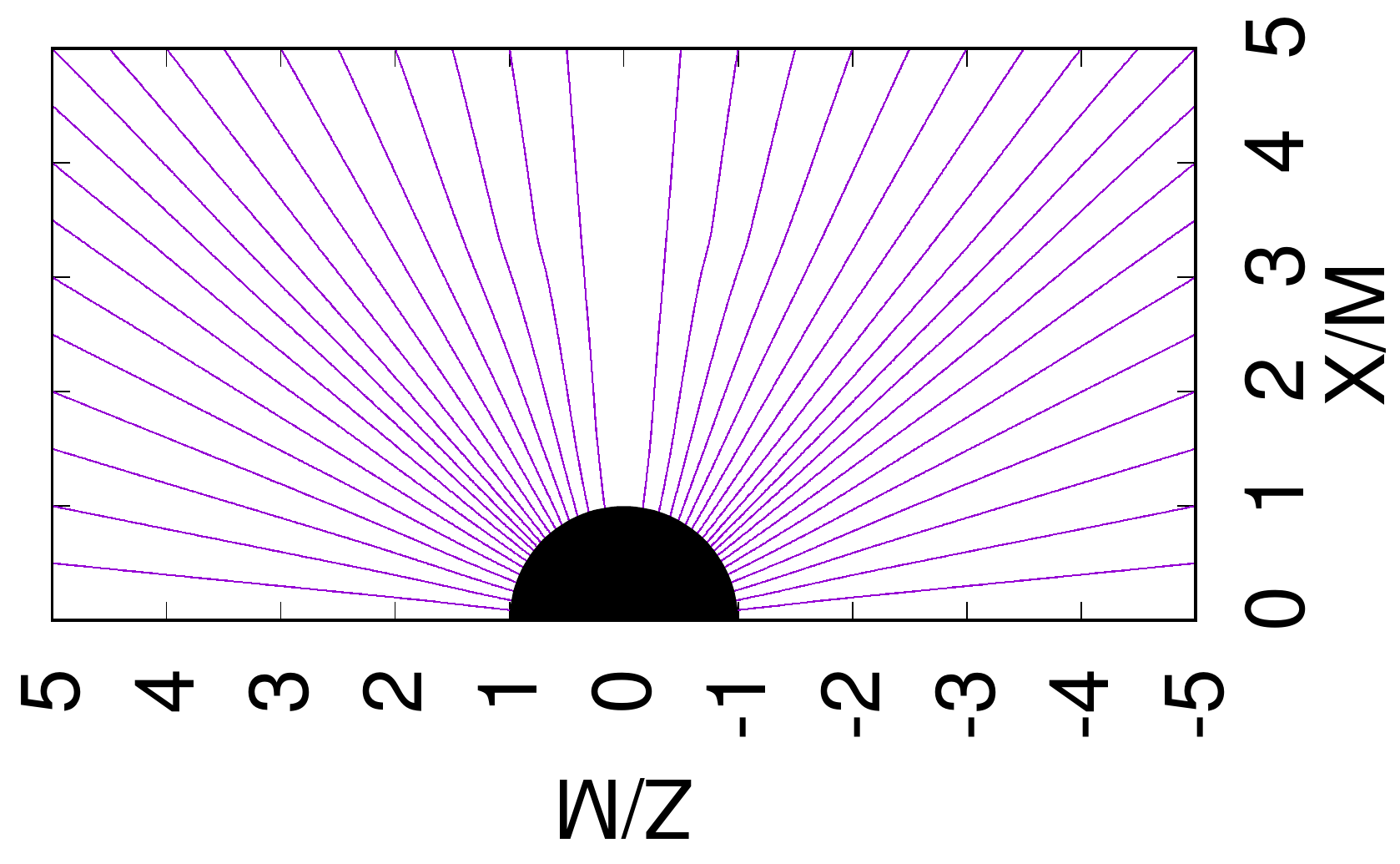}%
\label{fig:SplitMonopolec}
}%
\caption{Split Monopole: Magnetic field structure at a) $t=0M$, 
  b) $t=5M$ (without current sheet prescription), 
  and c) $t=5M$ (with current sheet prescription). 
  Formatting for b) and c) duplicates that of
  Refs.~\cite{Komissarov:2004,Paschalidis:2013} and \cite{McKinney:2006} respectively 
  so that direct comparison can be made.}
\label{fig:SplitMonopole}
\end{figure*}

The results of this test without a current sheet prescription are
shown in Fig.~\ref{fig:SplitMonopoleb}, and are in good agreement with
the ones obtained in Refs.~\cite{Komissarov:2004,Paschalidis:2013}.
If the prescription for handling equatorial current sheets is enabled,
no magnetic reconnection appears, as shown in
Fig.~\ref{fig:SplitMonopolec}, which agrees well with Fig.~4 of
Ref.~\cite{McKinney:2006}.

\subsubsection{Exact Wald}

This solution to Maxwell's equations, found by Wald~\cite{Wald:1974},
describes the electrovacuum around a rotating black hole immersed in a uniform magnetic field aligned with the axis of rotation of the black hole. 
Near the black hole, the solution contains both magnetic and electric fields, while far away the electric field vanishes. For a nonspinning
($a=0$) black hole, this is also a force-free solution, which is the
case considered in this test.   
Following Ref.~\cite{Paschalidis:2013},
we choose Kerr-Schild coordinates with a radial shift $r_0=0.4 M$
which yields a BH horizon located at $r=1.6M$ (see Appendix for
full 3+1 form of shifted Kerr-Schild spacetime).

Demonstrating convergence of the numerical to the exact solution in
this case is complicated by the large region inside the horizon that
violates the force-free condition $B^2-E^2 > 0$. To wit, in
Kerr-Schild coordinates Wald's solution yields
\beq
B^2 - E^2 = B_0^2 \left( 1-\frac{2 M \sin^2(\theta)}{r}\right).
\eeq
The horizon exists at $r=2M$ in these (unshifted Kerr-Schild)
coordinates, so in the equatorial plane ($\theta=\pi/2$), $B^2-E^2 >
0$ is violated at all points $r\le 2M$ - i.e., all points inside the
horizon, including the horizon itself. This has been noted previously
by Komissarov~\cite{Komissarov:2004}.

As such, when we apply the GRFFE constraints (as described in Step 4 of
Sec.~\ref{algorithms}) immediately after the initial data
are set up at $t=0$, all points where $B^2-E^2 \le 0$ {\it get overwritten to a
solution inconsistent with the Wald solution}. Hence we have a 
{\it non-stationary} solution inside the horizon. But despite this abrupt
replacement of data inside the horizon,
Fig.~\ref{fig:Exact_and_MagnetosphericWald} demonstrates the magnetic
field lines at the initial time overlap the lines at $t=5M$ extremely
well, indicating excellent {\it qualitative} agreement of our
numerical evolution with the stationarity of Wald's solution.

However, {\it quantitative} convergence of the numerical solution to
the stationary (initial) solution near the horizon is strongly
influenced by our code's correction of the $B^2-E^2 > 0$
violation inside and at the horizon, which induces spurious,
numerically-driven dynamics inside the horizon. Since numerical errors
can propagate superluminally, these dynamics propagate outside the
horizon and manifest as a numerical solution outside the
horizon {\it inconsistent} with stationarity and result in a drop in
the convergence order to the stationary solution within this region.

Therefore, to properly measure the order at which our numerical solution
converges to the stationary solution, we must ignore the
non-stationary region very close to the horizon. Our measurement
strategy is as follows. For a given numerically-evolved quantity $Q$,
we compute the $L_2$ norm of the difference between $Q$ and
its stationary value $Q_0$ (given by the Wald solution) as
\beq
\label{L2errordef}
\Delta Q = \sqrt{\int_{\mathcal{V}}\left(Q - Q_0\right)^2 d^3x}.
\eeq
We choose a volume $\mathcal{V}$ that covers the entire numerical
simulation domain, excising the region $r<8r_{\rm H}$, where 
$r_{\rm H}$ is the horizon radius in our shifted Kerr-Schild
coordinates ($r_{\rm H}\approx 1.4M$). Additionally, to eliminate
known low-order convergence contamination from the chosen approximate
outer boundary conditions, the region $r>90M$ is also excised.

Table~\ref{tab:ewaldconverge} demonstrates
that when adopting this measure, our numerical results converge to the
stationary (initial) solution at approximately second order in
numerical grid spacing (i.e., the error $E$ is measured to scale as
$E\sim{\Delta x_{\rm min}}^2$), which is expected from our choice
of reconstruction scheme and AMR grid interpolation order.

\begin{table}
\footnotesize
\centering
\begin{tabular}{ccccc}
\hline\hline
Res. & $\Delta A^x\cdot10^4$ & $\Delta A^y\cdot10^4$ & $\Delta A^z\cdot10^5$ & $\Delta(\psi^6\Phi)\cdot10^4$ \\ [0.5ex]
\hline

H & $4.855$ & $4.857$ & $4.394$ & $1.053$ \\

M & $3.9\cdot$H & $3.9\cdot$H & $4.3\cdot$H & $4.0\cdot$H \\

L & $15.3\cdot$H & $15.3\cdot$H & $19.3\cdot$H & $15.6\cdot$H \\ 
\end{tabular}
\begin{tabular}{ccccccc}

\hline\hline
Res. & $\Delta v^x\cdot10^4$ & $\Delta v^y\cdot10^4$ & $\Delta v^z\cdot10^5$ & $\Delta B^x\cdot10^4$ & $\Delta B^y\cdot10^4$ & $\Delta B^z\cdot10^4$ \\ [0.5ex]
\hline

H & $2.115$ & $2.102$ & $1.209$ & $1.406$ & $1.398$ & $8.764$ \\

M & $3.3\cdot$H & $3.4\cdot$H & $3.5\cdot$H & $3.6\cdot$H & $3.7\cdot$H & $4.0\cdot$H \\

L & $12.1\cdot$H & $12.2\cdot$H & $13.1\cdot$H & $14.0\cdot$H & $14.1\cdot$H & $15.8\cdot$H \\ 
\hline
\end{tabular}
\caption{$L_2$ norms of the difference between our numerical results
  and the initial data (Eq.~\ref{L2errordef}) at $t=5M$ for the
  4-vector potential (top), and the velocity and magnetic field
  (bottom) at 3 resolutions. L, M, and H represent choice of numerical
  resolution, as defined in Table~\ref{tab:3Dgrid}.}
\label{tab:ewaldconverge}
\end{table}

\subsubsection{Magnetospheric Wald}

The magnetospheric Wald problem, called the ``ultimate Rosetta
Stone'' by Komissarov \cite{Komissarov:2004}, yields
insights about black hole magnetospheres beyond the Membrane Paradigm
\cite{Thorne:1986book}. The Membrane Paradigm predicts that only a
small fraction of magnetic field lines penetrating a spinning black
hole's horizon will be dragged via Lens-Thirring into a co-rotating
motion with the BH spin. The magnetospheric Wald problem is a clear
counterexample to this prediction. 

\begin{figure*}
\centering
\subfloat[]{%
\includegraphics[angle=0,scale=0.85]{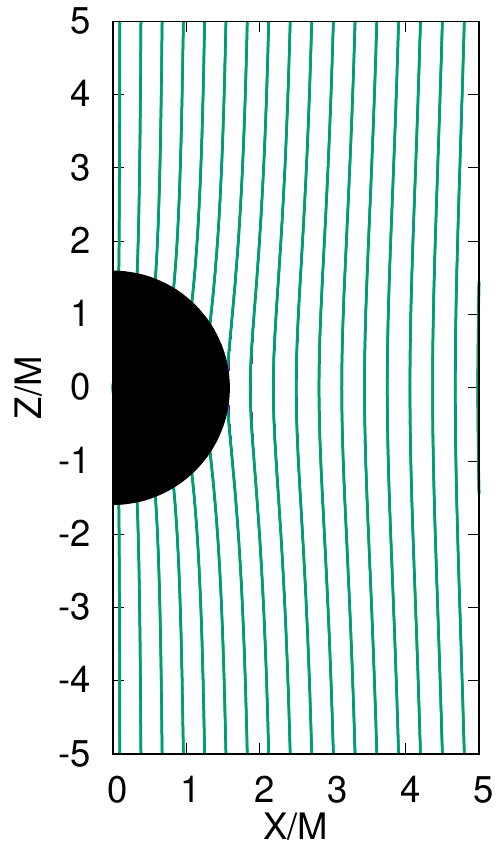}%
}
\quad
\subfloat[]{%
\includegraphics[angle=0,scale=0.6]{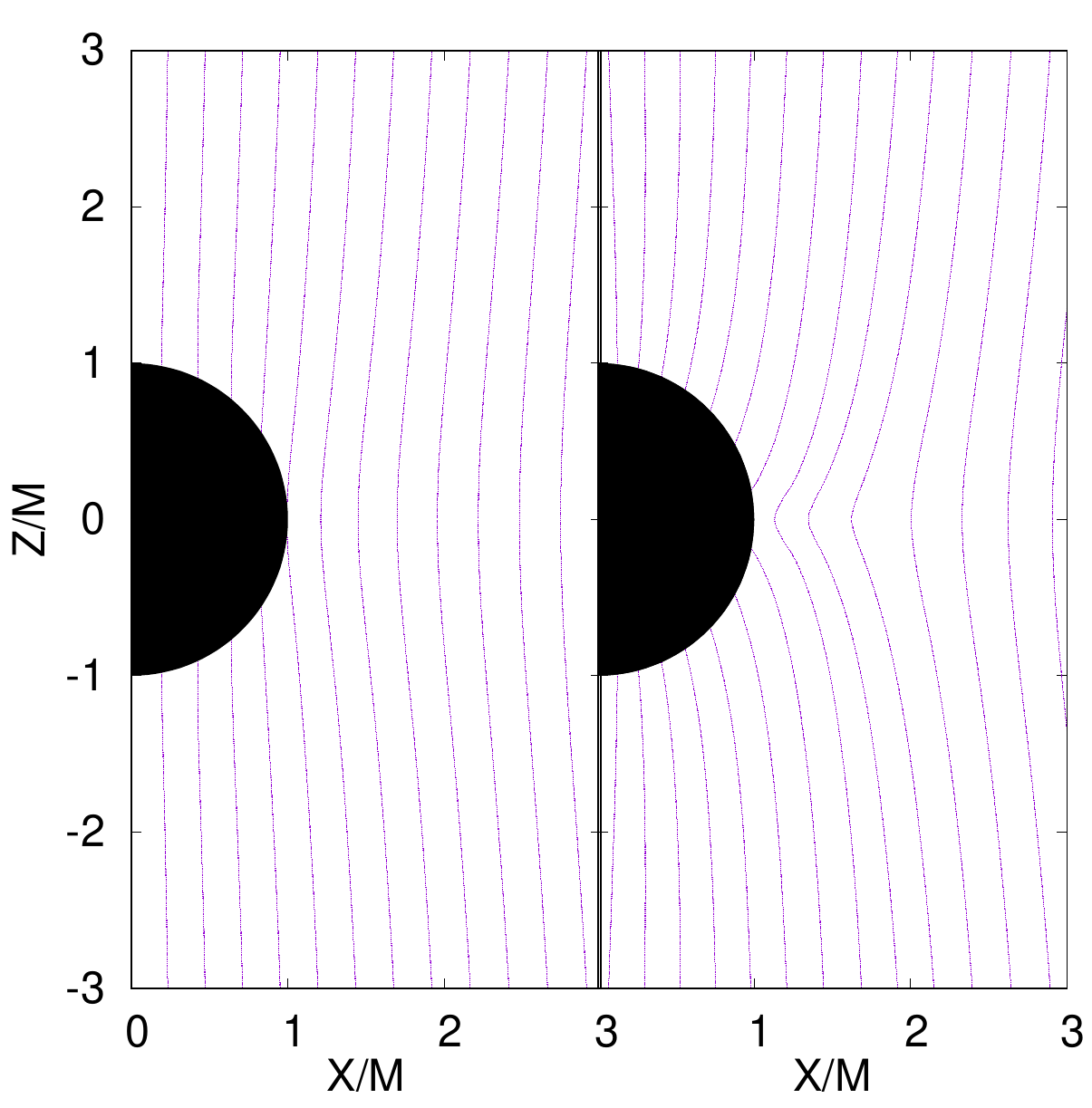}%
}%
\caption{%
a) Exact Wald: Magnetic field structure at $t=5M$ (green) plotted atop exact solution (i.e., $t=0$ data; purple).
b) Magnetospheric Wald: Magnetic field structure at $t=0M$ (left) and after reaching apparent equilibrium at $t=126M$ (right). 
Figure formatting duplicates that of Ref.~\cite{Paschalidis:2013} so that direct comparisons can be made.}
\label{fig:Exact_and_MagnetosphericWald}%
\end{figure*}

There is no known analytic solution to this problem as $t\to\infty$,
but the initial data are expected to evolve into a steady,
equatorially-symmetric state with a current sheet visible in the
equatorial plane within the black hole ergosphere. 
The right panel of Fig.~\ref{fig:Exact_and_MagnetosphericWald} shows
the poloidal magnetic field lines at $t = 126M$, where the solution
reaches a steady state, which agrees well with the state found in
Refs.~\cite{Komissarov:2004,Paschalidis:2013}.
As in Ref.~\cite{Paschalidis:2013}, this
test is performed in shifted Kerr-Schild radial coordinates with
dimensionless spin $a_* = 0.9$ and radial shift $r_0=0.4359 M$
selected so that the horizon radius is nearly one.

\section{Conclusions and Future Work}
\label{conclusions}

We have presented code validation test results from our new code,
\GiR, which is the first open-source GRFFE code designed to model
strongly-magnetized, tenuous plasmas in full general relativity. The
GRFFE approximation is well motivated in many
cases of astrophysical interest, including the launching of
relativistic jets in BH and NS spacetimes, and the evolution of NS,
pulsar, and BH magnetospheres both in isolation and when interacting
with binary companions.  

To validate this new code, it was subjected to a battery of tests in
both flat and strongly-curved BH spacetimes. \GiR successfully
reproduces several classes of exact smooth and discontinuous wave
solutions in one spatial dimension. We also demonstrated that for a
test case in which a transition layer evolves from a state that
satisfies the FFE conditions to one that violates them, \GiR behaves
in the expected manner. Moving from one to three spatial dimensions,
we also verified that our code reproduces the known steady-state
solution for the aligned rotator, which is frequently used to
approximate pulsar magnetospheres.

A suite of three tests were performed in three spatial
dimensions within a shifted-radius, Kerr-Schild curved spacetime
background (as described in the Appendix), and endowed with a current
sheet in the equatorial plane. All three cases evolved to known
solutions, with the exact Wald test validating that our code
is, as expected, second-order convergent to the exact solution. By
passing this suite of tests, we have shown that \GiR passes all of
the validation tests adopted by the GRFFE codes of
Refs.~\cite{McKinney:2006,Paschalidis:2013}. \GiR has therefore
demonstrated its capacity for evolving force-free fields in dynamical
spacetimes.

At present, \GiR can perform purely force-free evolutions in the
context of dynamical spacetime evolutions within the Einstein
Toolkit/Carpet/McLachlan framework. In the future, we
plan to integrate \GiR with \IGM, so that it can tackle problems
that combine both GRMHD and GRFFE domains. This will involve the
design and implementation of new GRMHD/GRFFE matching algorithms, and
will likely benefit from the creation of a separate library
containing common features of \GiR and \IGM, to avoid unnecessary code
duplication. As outlined throughout the text, there are a number
of problems of astrophysical interest that \GiR is ideally positioned
to explore, and we hope the wider community will join us in these
investigations by leveraging this new open-source tool.

\section*{Acknowledgements}
We gratefully acknowledge I. Ruchlin and V. Paschalidis for valuable
discussions as this work was prepared, as well as N. Gregg for
technical assistance. This work was supported by NASA Grant
13-ATP13-0077 and NSF EPSCoR Grant OIA-1458952. Large-scale computations
were performed on West Virginia University's {\tt Spruce Knob}
supercomputer, funded in part by NSF EPSCoR Research Infrastructure
Improvement Cooperative Agreement \#1003907, the state of West
Virginia (WVEPSCoR via the Higher Education Policy Commission), and
West Virginia University. MBW wishes to acknowledge H.-I. Kim for technical assistance
with part of the computations done in Korea for this work. 

\appendix
\section*{Appendix: 3+1 Black Hole Spacetime in Shifted Kerr-Schild
  Coordinates}
\label{app:shiftedKS}  

\setcounter{section}{1}
A complete description of a black hole spacetime in Kerr-Schild
spherical polar coordinates that includes an explicit analytic form of
the extrinsic curvature for arbitrary spin parameters does
not exist in the literature, so we first include it here for
completeness. Then we present our strategy for transforming spacetime
quantities into shifted Kerr-Schild Cartesian coordinates.

In unshifted spherical polar coordinates, where $\rho=r^2+a^2 \cos^2(\theta)$,
$M$ is the black hole mass, and $a$ is the black hole spin parameter,
the Kerr-Schild lapse, shift, and 3-metric are given by
\beqn
\alpha   &=& \frac{1}{\sqrt{ 1 + \frac{2 M r}{\rho^2} }} \\
\beta^r  &=& \alpha^2\frac{2 M r}{\rho^2} \\
\beta^\theta &=& \beta^\phi = \gamma_{r\theta}=\gamma_{\theta\phi}= 0 \\
\gamma_{rr}    &=& 1 + \frac{2 M r}{\rho^2}\\
\gamma_{r\phi}   &=&-a \gamma_{rr} \sin^2(\theta)\\
\gamma_{\theta\theta}  &=& \rho^2\\
\gamma_{\phi\phi}  &=& \left( r^2 + a^2 + \frac{2 M r}{\rho^2} a^2 \sin^2(\theta) \right) \sin^2(\theta).
\eeqn

Next, we define a few useful quantities,
\beqn
A&=&\left(a^2 \cos (2 \theta )+a^2+2 r^2\right) \\
B&=&A+4Mr \\
D&=&\sqrt{\frac{2 M r}{a^2 \cos ^2(\theta)+r^2}+1}.
\eeqn
Then the extrinsic curvature 
$K_{ij}=(\nabla_i \beta_j +\nabla_j \beta_i)/(2\alpha)$ 
(see, \textit{e.g.}, Eq. 13 in Ref.~\cite{Cook2000}) with $\partial_t
\gamma_{ij}=0$, may be written in spherical polar coordinates as
\beqn
\fl K_{rr}&=&\frac{D (A+2 M r)}{A^2 B}\left[ 4 M \left(a^2 \cos (2 \theta )+a^2-2 r^2\right)\right] \\
\fl K_{r\theta}&=&\frac{D}{AB}\left[ 8 a^2 M r \sin (\theta ) \cos (\theta )\right] \\
\fl K_{r\phi}&=&\frac{D}{A^2}\left[ -2 a M \sin ^2(\theta ) \left(a^2 \cos (2 \theta )+a^2-2 r^2\right)\right] \\
\fl K_{\theta\theta}&=&\frac{D}{B}\left[ 4 M r^2 \right]\\
\fl K_{\theta\phi}&=&\frac{D}{AB}\left[ -8 a^3 M r \sin ^3(\theta ) \cos (\theta )\right] \\
\fl K_{\phi\phi}&=&\frac{D}{A^2 B}\left[ 2 M r \sin ^2(\theta )
  \left(a^4 (r-M) \cos (4 \theta )+a^4 (M+3 r)+\right.\right. \\
\fl && \left.\left.\quad\quad\quad\quad\quad\quad\quad 4 a^2 r^2 (2 r-M)+4 a^2 r \cos (2 \theta ) \left(a^2+r (M+2 r)\right)+8 r^5\right)\right].
\eeqn

All \GiR curved-spacetime code validation tests adopt shifted
Kerr-Schild Cartesian coordinates $(x',y',z')$, which map
$(0,0,0)$ to the finite radius $r=r_0>0$ in standard (unshifted)
Kerr-Schild spherical polar coordinates. So, in many ways, this is similar
to a trumpet spacetime. Though this radial shift acts to shrink the
black hole's coordinate size, it also renders the very
strongly-curved spacetime fields at $r<r_0$ to vanish deep inside the horizon,
which can contribute to numerical stability when evolving
hydrodynamic, MHD, and FFE fields inside the horizon.

The shifted radial coordinate $r'$ relates to the standard spherical
polar radial coordinate $r$ via $r=r'+r_0$, where $r_0>0$ is the
(constant) radial shift. 

As an example, to compute $K_{x'y'}$ at some arbitrary point
$(x',y',z')$, we first convert the coordinate $(x',y',z')$ into shifted
spherical-polar coordinates via
$(r'=\sqrt{x'^2+y'^2+z'^2},\theta',\phi')=(r',\theta,\phi)$, as a purely
radial shift like this preserves the original angles. Next, we
evaluate the components of the Kerr-Schild extrinsic curvature
$K_{ij}$ (provided above) in standard spherical polar coordinates at
$(r=r'+r_0,\theta,\phi)$. Defining $x^i_{\rm sph,sh}$ as the $i$th
shifted spherical polar coordinate and $x^i_{\rm sph}$ as the $i$th
(unshifted) spherical polar coordinate, $K_{x'y'}$ is computed via the
standard coordinate transformations:
\beq
K_{x'y'} = \frac{dx^k_{\rm sph,sh}}{dx'} \frac{dx^l_{\rm sph,sh}}{dy'} 
\frac{dx^i_{\rm sph}}{dx^k_{\rm sph,sh}} \frac{dx^i_{\rm sph}}{dx^l_{\rm sph,sh}} K_{ij}. 
\eeq
However, we have $dx^i_{\rm sph} = dx^i_{\rm sph,sh}$, since the radial shift $r_0$
is a constant and the angles are unaffected by the radial shift. This
implies that $dx^k_{\rm sph,sh}/dx' = dx^k_{\rm sph}/dx'$ and 
$dx^i_{\rm sph}/dx^k_{\rm sph,sh}=\delta^i_k$.

So after computing {\it any spacetime quantity} at a
point $(r=r'+r_0,\theta,\phi)$, we need only apply the standard
spherical-to-Cartesian coordinate transformation to evaluate that
quantity in shifted Cartesian coordinates.

\section*{References}

\bibliographystyle{plain}
\bibliography{references}

\end{document}